\documentclass[twoside,twocolumn,9pt]{article}
\usepackage{extsizes}
\usepackage[super,sort&compress,comma]{natbib} 
\usepackage[version=3]{mhchem}
\usepackage[left=1.5cm, right=1.5cm, top=1.785cm, bottom=2.0cm]{geometry}
\usepackage{balance}
\usepackage{mathptmx}
\usepackage{sectsty}
\usepackage{graphicx} 
\usepackage{lastpage}
\usepackage[format=plain,justification=justified,singlelinecheck=false,font={stretch=1.125,small,sf},labelfont=bf,labelsep=space]{caption}
\usepackage{float}
\usepackage{fancyhdr}
\usepackage{fnpos}
\usepackage[english]{babel}
\addto{\captionsenglish}{%
  
}
\usepackage{array}
\usepackage{droidsans}
\usepackage{charter}
\usepackage[T1]{fontenc}
\usepackage[usenames,dvipsnames]{xcolor}
\usepackage{setspace}
\usepackage[compact]{titlesec}
\usepackage{hyperref}

\usepackage{epstopdf}

\definecolor{cream}{RGB}{222,217,201}

\begin{document}

\pagestyle{fancy}
\thispagestyle{plain}
\fancypagestyle{plain}{
\renewcommand{\headrulewidth}{0pt}
}

\makeFNbottom
\makeatletter
\renewcommand\LARGE{\@setfontsize\LARGE{15pt}{17}}
\renewcommand\Large{\@setfontsize\Large{12pt}{14}}
\renewcommand\large{\@setfontsize\large{10pt}{12}}
\renewcommand\footnotesize{\@setfontsize\footnotesize{7pt}{10}}
\makeatother

\renewcommand{\thefootnote}{\fnsymbol{footnote}}
\renewcommand\footnoterule{\vspace*{1pt}%
\color{cream}\hrule width 3.5in height 0.4pt \color{black}\vspace*{5pt}} 
\setcounter{secnumdepth}{5}

\makeatletter 
\renewcommand\@biblabel[1]{#1}            
\renewcommand\@makefntext[1]%
{\noindent\makebox[0pt][r]{\@thefnmark\,}#1}
\makeatother 
\renewcommand{\figurename}{\small{Fig.}~}
\sectionfont{\sffamily\Large}
\subsectionfont{\normalsize}
\subsubsectionfont{\bf}
\setstretch{1.125} 
\setlength{\skip\footins}{0.8cm}
\setlength{\footnotesep}{0.25cm}
\setlength{\jot}{10pt}
\titlespacing*{\section}{0pt}{4pt}{4pt}
\titlespacing*{\subsection}{0pt}{15pt}{1pt}

\fancyfoot{}
\fancyfoot[LO,RE]{\vspace{-7.1pt}\includegraphics[height=9pt]{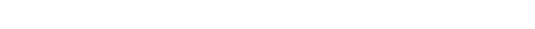}}
\fancyfoot[CO]{\vspace{-7.1pt}\hspace{13.2cm}\includegraphics{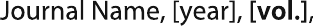}}
\fancyfoot[CE]{\vspace{-7.2pt}\hspace{-14.2cm}\includegraphics{head_foot/RF}}
\fancyfoot[RO]{\footnotesize{\sffamily{1--\pageref{LastPage} ~\textbar  \hspace{2pt}\thepage}}}
\fancyfoot[LE]{\footnotesize{\sffamily{\thepage~\textbar\hspace{3.45cm} 1--\pageref{LastPage}}}}
\fancyhead{}
\renewcommand{\headrulewidth}{0pt} 
\renewcommand{\footrulewidth}{0pt}
\setlength{\arrayrulewidth}{1pt}
\setlength{\columnsep}{6.5mm}
\setlength\bibsep{1pt}

\makeatletter 
\newlength{\figrulesep} 
\setlength{\figrulesep}{0.5\textfloatsep} 

\newcommand{\topfigrule}{\vspace*{-1pt}%
\noindent{\color{cream}\rule[-\figrulesep]{\columnwidth}{1.5pt}} }

\newcommand{\botfigrule}{\vspace*{-2pt}%
\noindent{\color{cream}\rule[\figrulesep]{\columnwidth}{1.5pt}} }

\newcommand{\dblfigrule}{\vspace*{-1pt}%
\noindent{\color{cream}\rule[-\figrulesep]{\textwidth}{1.5pt}} }

\makeatother

\twocolumn[
  \begin{@twocolumnfalse}
{\includegraphics[height=30pt]{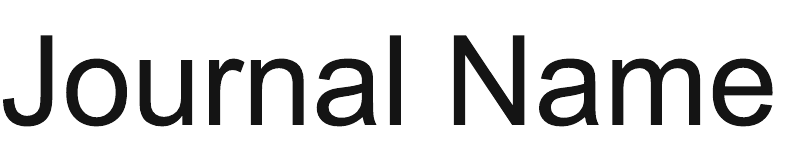}\hfill\raisebox{0pt}[0pt][0pt]{\includegraphics[height=55pt]{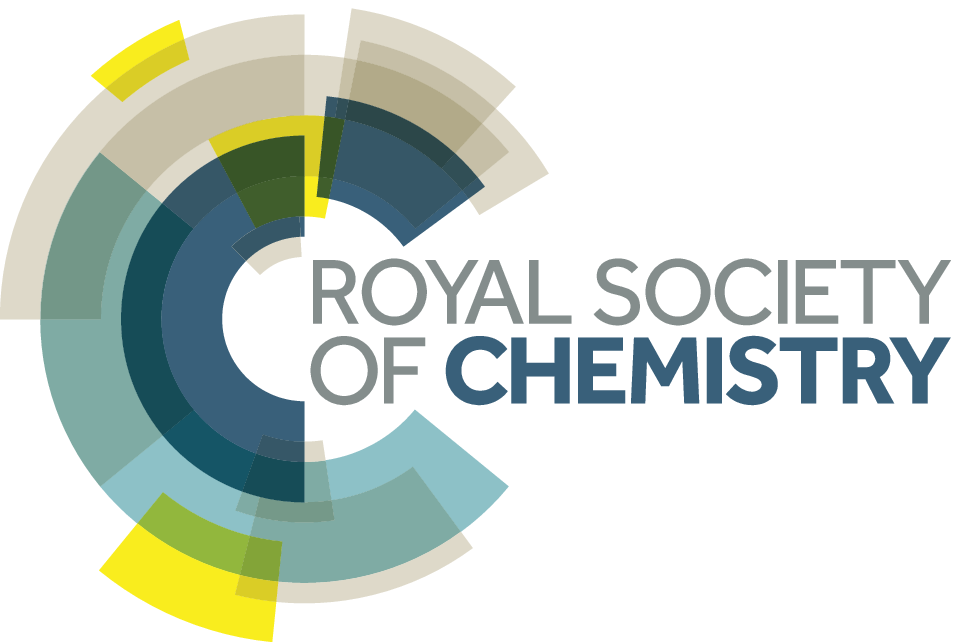}}\\[1ex]
\includegraphics[width=18.5cm]{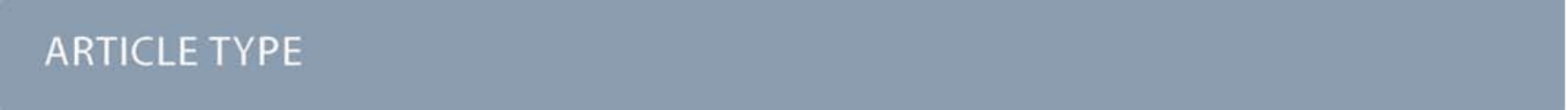}}\par
\vspace{1em}
\sffamily
\begin{tabular}{m{4.5cm} p{13.5cm} }

\includegraphics{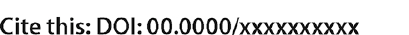} & \noindent\LARGE{\textbf{Self-assembly and percolation in magnetic colloids}} \\
\vspace{0.3cm} & \vspace{0.3cm} \\

 & \noindent\large{Hauke Carstensen,\textit{$^{a}$} Anne Kr\"amer,\textit{$^{a}$} Vassilios Kapaklis,\textit{$^{a}$} and Max Wolff\textit{$^{a}$}} \\

\includegraphics{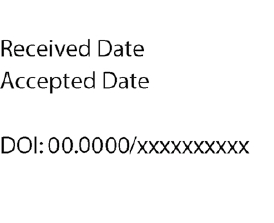} & \noindent\normalsize{We study the self-assembly of branching-chain networks and crystals in a binary colloidal system with tunable interactions. The particle positions are extracted from microscope images and order parameters are extracted by image processing and statistical analysis. With these, we construct phase diagrams with respect to particle density, ratio and interaction. In order to draw a more complete picture, we complement the experiments with computer simulations. We establish a region in the phase diagram, where bead ratios and interactions are symmetric, promoting percolated structures.} \\

\end{tabular}

 \end{@twocolumnfalse} \vspace{0.6cm}

  ]

\renewcommand*\rmdefault{bch}\normalfont\upshape
\rmfamily
\section*{}
\vspace{-1cm}


\footnotetext{\textit{$^{a}$~Department of Physics and Astronomy, Uppsala University, Box 516, SE-75120 Uppsala, Sweden; E-mail: max.wolff@physics.uu.se}}





\section{Introduction}

Self-assembly is the spontaneous formation of ordered structures on all length-scales~\cite{whitesides2002SCIENCE} and a new route for the design of materials with complex functionality~\cite{whitesides2002PNAS}. For small building blocks and finite temperatures entropy will allow a system to reach the global minimum in free energy and thermodynamic equilibrium. For systems dominated by the interaction between the constituents this may not be the case and the system can get trapped in a metastable state or become kinetically arrested~\cite{klokkenburg2006quantitative}.

Colloidal systems consist of microscopic beads solved in a carrier liquid. Typically, the size of these beads is large and thermal energy becomes negligible. The systems' phase behaviour is defined by the interactions between the beads, making them interesting for the study of phase formation and self-assembly. By coating the bead surface with polymers, soft steric interactions may be introduced~\cite{Steric_Langmuir_2007}. More long-ranged interactions that are possible to realize, are Coulomb interactions. The self-assembly of charged colloids was studied for beads of different size~\cite{C2SM26192H}, aspect ratio~\cite{demirors2010directed} as well as opposite charge~\cite{kitaev2003self}.
Even more rich in behaviour are magnetic interactions, since these are of dipolar nature and may be altered by external magnetic fields~\cite{du2013generating}. The anisotropy of the interaction leads to the formation of clusters in form of rings or chains with a head to tail orientation of the individual particles \cite{Kantorovich2008}. With decreasing temperature or increasing density the size of such clusters grows and networks may form. These are challenging to study for nanometer sized particles as they require frozen samples to allow studies with transmission electron microscopy \cite{Butter2003} or computer simulations \cite{PhysRevLett.71.2729, Kantorovich2015}. For larger particles the assembly process is easier to follow \cite{PhysRevE.59.R4758} and rings, chains and small networks were reported. Illustrative in this context is the study of cm sized particles with different magnetic moments and their pattern formation and segregation \cite{PhysRevE.68.026207, PhysRevE.70.031304}. It was shown that by shaking and quenching magnetic particle systems larger open networks may be formed \cite{PhysRevE.67.021302, Kroger2018}.
All the above mentioned studies where done without externally applied magnetic field. Time dependent magnetic fields, however, allow precise tuning of the interactions~\cite{Pham2017} and  control of the self-assembly process~\cite{alert2014landscape} as well as the study of the kinetics of self-assembly~\cite{C4SM00132J} and even the motion of self-assembled magnetic surface swimmers~\cite{snezhko2009self}.
By dispersing the magnetic beads in a ferrofluid matrix, an effective moment may be introduced, leading to tunable particle interactions. The local self-assembly of such systems has been studied~\cite{erb2009magnetic} and a phase diagram of the local coordination was presented~\cite{khalil2012binary}. For a two dimensional systems with in plane magnetic field and depending on the interaction between the beads cubic or hexagonal phases may form~\cite{carstensen2015phase}.

In technology, magnetic liquids are applied, for example, in loud speakers~\cite{rosensweig2008study}, tunable photonic \cite{photonic, he2012magnetic, saado2002tunable} or plasmonic structures~\cite{fan2013magnetically} and tunable shock absorbers~\cite{nguyen2009optimal}. These applications require knowledge on phase formation, not only on the local length-scales but also global, e.g. the formation of clusters or extended networks. On intermediate length-scales colloidal crystals or networks may form, which for micrometer-sized particles allow tuning of the viscosity in so called magneto-rheological materials~\cite{C0SM00014K, wang2011magnetorheological}.

An important concept in this context is jamming, which allows relating changes in macroscopic material properties, e.g. viscosity, to topological constrains~\cite{ohern2003PRE}. In the case of attractive interactions a phase diagram of jamming has been established~\cite{trappe2001NATURE} and it was shown that the jamming transition in systems with attractive and repulsive interactions, fall into different universality classes~\cite{Charles_2005_PRL, koeze2018PRL}. Similar to jamming, percolation, may affect the macroscopic properties of a system. In the percolated state  all parts of a system are connected with each other. The jamming as well as percolation transition are characterised by critical densities~\cite{achlioptas2009SCIENCE}.
The critical density for percolation was calculated for different coordinations in two and three dimensions~\cite{scher1970JCP} and is very similar to jamming happening at a critical random-close-packing density~\cite{ohern2003PRE}. The influence of a long range repulsion and short range attraction on the percolation transition was investigated systematically by Monte-Carlo simulations~\cite{valadez2013RSC}. These show that long-ranged repulsion increases the average cluster size and lowers the volume fraction for the percolation threshold.

In this communication, we report on the self-assembly and percolation of a binary, two dimensional system of magnetic beads with tunable interactions in an out-of-plane magnetic field. We classify the cluster formation with respect to the coordination of the individual beads, the crystal symmetry and percolation. We find that the critical density for percolation depends on the interaction between the beads as well as the bead ratio and report the lowest critical density for a bead ratio of 1:1 with attraction between the two species of beads and repulsion among beads of the same type.

\section{Experimental details}

Monodisperse micrometer-sized spherical particles (beads) -- one magnetic and one non-magnetic -- are suspended in a water-based ferrofluid. The beads, purchased from Microparticles GmbH, are composed of polystyrene and have a diameter of 10~$\mu$m. The magnetic beads are coated by a shell of magnetic nano-particles. The ferrofluid (FF) consists of 10~nm iron oxide (Fe$_3$O$_4$) nanoparticles, dispersed in water and purchased from LiquidResearch. In suspension, the beads exhibit an apparent magnetic behaviour mediated by the surrounding ferrofluid~ \cite{skjeltorp1984crystallization, carstensen2015phase, carstensen2018statistical}. The effective magnetic susceptibility of the suspended beads is modified by the susceptibility of the ferrofluid. This effect is in analogy to the concept of effective densities for objects in fluids, as described by the Archimedes principle, resulting in an apparent reduction of the susceptibility of the magnetic beads. The non-magnetic beads form magnetic voids in the ferrofluid matrix and may  be assigned an effective magnetic susceptibility. By changing the concentration of magnetic nano-particles in the ferrofluid, the effective moments and thus the interaction between the beads can be tuned.

\begin{figure}[t!]
	\centering		
	\includegraphics[width=0.9\columnwidth]{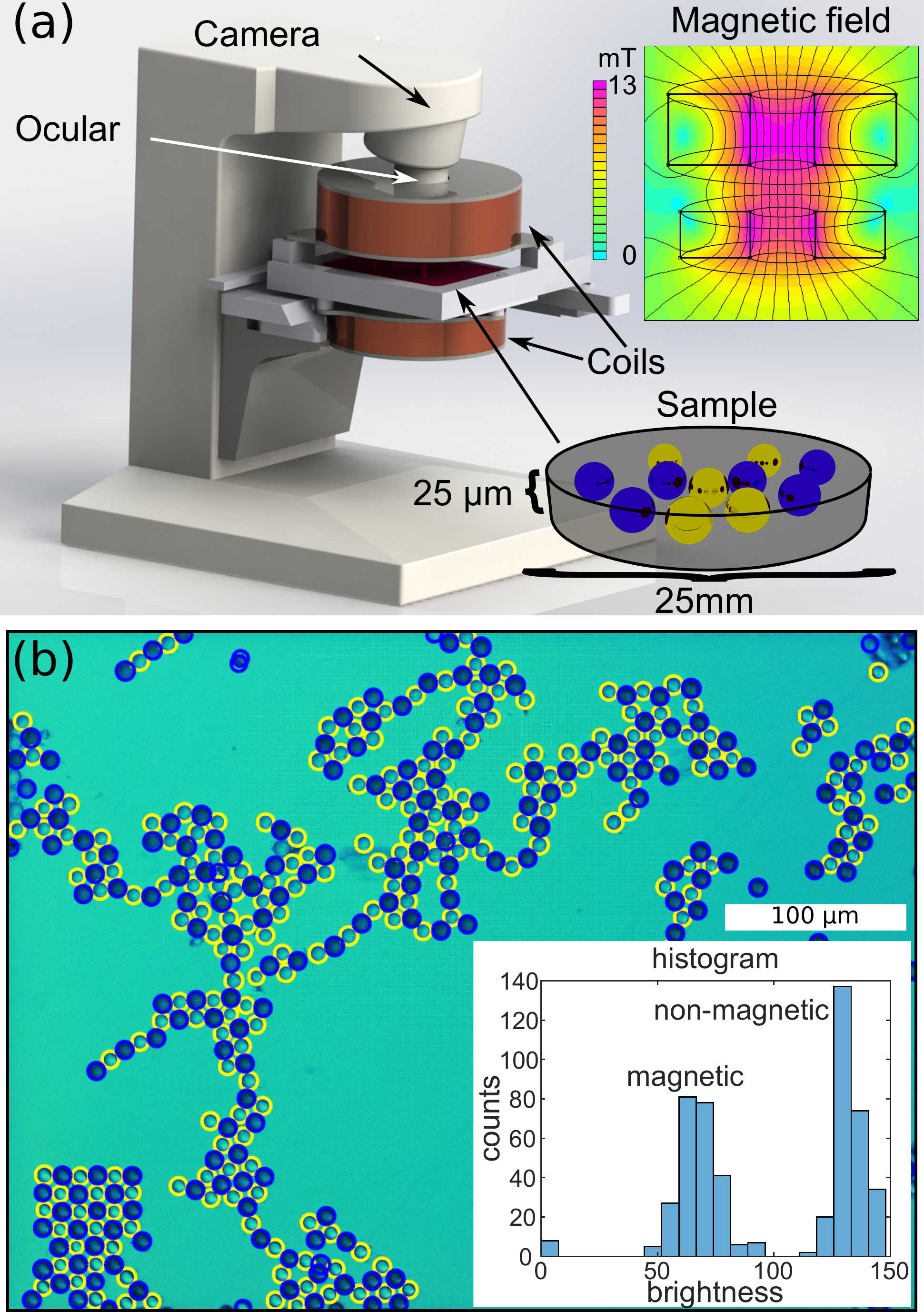}
	\caption{Panel (a) depicts the transmission microscope setup. The panel to the upper right shows the magnetic flux density. Panel (b) is a typical microscope image with the magnetic and non-magnetic beads in darker and brighter color, respectively. The particles have been identified automatically by evaluating their brightness (bottom right).}
	\label{setup}
\end{figure}

In the limit, $\chi_\text{M}<<1$ and $\chi_\text{FF}<<1$, with $\chi_\text{M}$ and $\chi_\text{FF}$ being the susceptibilities of the magnetic beads and the ferro-fluid, respectively, the effective bead susceptibilities $\chi_\text{M,eff}$ and $\chi_\text{N,eff}$, for the magnetic and non-magnetic beads, respectively, are~\cite{khalil2012binary, erb2009magnetic, carstensen2015phase, phdthesisHauke}:
\begin{align}
\chi_\text{M,eff}&\propto \chi_\text{M}-\chi_\text{FF}\\
\chi_\text{N,eff}&\propto -\chi_\text{FF}
\end{align}
If the beads are confined along one dimension and a magnetic field is applied along the confinement direction, the interaction between the beads can be classified as follows:
\begin{itemize}
    \item The interaction between identical beads is repulsive but tunable in strength, by changing the concentration of magnetic nano-particles in the ferrofluid.
    \item For $\chi_\text{M,eff}=0$ and $\chi_\text{N,eff}=0$ the magnetic and non-magnetic beads, respectively, are non-interacting.
\end{itemize}
The interaction between the two types of beads can be further modified in the following ways \cite{carstensen2015phase}: 
\begin{itemize}
    \item For $\chi_\text{N,eff} < 0 < \chi_\text{M,eff}$ the coupling is anti-ferromagnetic and an attractive force is present between magnetic and non-magnetic beads.
    \item For $\chi_\text{M,eff} < 0$ and $\chi_\text{N,eff} < 0$ the interaction between all particles is repulsive.
\end{itemize}
This tunable interaction allows systematical study of the structure formation in this binary system, with the ground states in the local ordering having been established previously~ \cite{khalil2012binary,yang2013tunable,carstensen2015phase,carstensen2018statistical}. Here, we focus on the structures formation on larger length-scales and percolation phenomena.

The samples were prepared by mixing 20~$\mu l$ ferrofluid, 5~$\mu l$ (1\% w/w) magnetic beads dispersion and 1~$\mu l$ (2.5\% w/w) non-magnetic beads dispersion. This results in approximately equal numbers of magnetic and non-magnetic beads present in the sample. However, the beads may be inhomogeneously distributed and the local densities and ratios vary spatially within each sample. This enables systematical study of the effect of density and bead ratio on the self-assembly. To vary the interaction between the microbeads, a series of samples with varying FF concentration were prepared. All of the samples were confined between two glass slides separated by an oil-covered 25~$\mu$m thick Mylar ring and studied by transmission optical microscopy. An out-of-plane magnetic field, with a maximum strength of 14~mT, was applied using a pair of Helmholtz coils. The magnetic flux density was calculated with finite element methods, implemented in FEMM \cite{FEMM, carstensen2015phase}. Fig.~\ref{setup}(a) shows a sketch of the experimental setup and the upper right panel results of the field calculation, indicating that the magnetic field is homogeneous over the area of the sample. The sample was loaded into the sample cell at zero field and all microscope images were taken at an out-of-plane field of 0.14 mT. After loading the samples and before acquiring the images with the microscope, the field is cycled (switched on and off) repeatedly in order to approach an equilibrium structure. Subsequently, microscope images are recorded and bead positions are automatically extracted, identifying circlular structures using image processing routines in MATLAB\textsuperscript{\textregistered}. The bead types are distinguished by their apparent brightness in the images. In addition, the automatic detection was made more robust, utilizing a neural network algorithm trained on three images with labeled data (bead positions and types).

        \begin{figure*}[t]
	        \centering		
	        \includegraphics[width=1\textwidth]{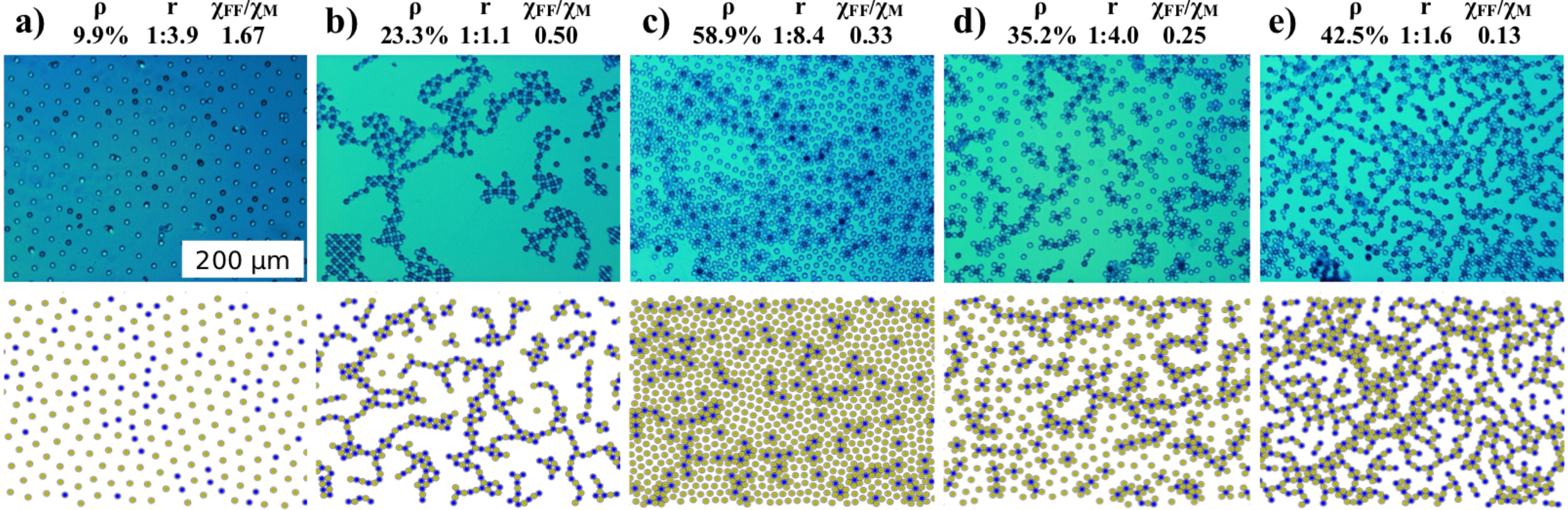}
	            \caption{Microscope images (top row) and simulated data (bottom row) for varying bead densities $\rho$, bead ratios $r$, and interactions $\chi_\text{FF}/\chi_\text{M}$.}
	        \label{fig:imagetable_comp15k}
            \end{figure*}

To complement the experiments, we performed computer simulations. Magnetic and non-magnetic beads are placed sequentially at random positions (without overlaps). The size of the simulated rectangular area was chosen to match the size of a typical microscope image ($50 \times 37$ bead diameters) and periodic boundary conditions were applied. We further assume zero temperature, justified by the large size of the beads and the negligible Brownian motion as well as significant viscosity of the ferrofluid (zero bead momentum), no inertia and an overdamped equation of motion. The beads' magnetic moment is described as out-of-plane dipoles. To avoid overlapping of particles a hard sphere potential was used and particles are placed next to each other once they touch. In each simulation step the bead displacements are calculated from the force acting upon each bead, resulting from the presence of all other beads, according to:

\begin{equation}
F_{j}= m_j c \sum_k \frac{m_k}{r_{j,k}^{-4}}
\end{equation}
where $m_j$ and $m_k$ are the dipole moments of beads separated by a distance $r_{j,k}$. As the dipolar forces decay rapidly with distance, we calculate the forces applying a cutoff of six bead diameters. Furthermore, the dipole force is scaled by $c=0.01$ to keep the displacements in each step much smaller then one bead diameter. Our simulations minimise the total energy of the system and typically converge after 15000 iteration steps.

\section{Results}

Representative microscope images are shown in Fig.~\ref{fig:imagetable_comp15k} (top row). The bead density $\rho$ defined as the image area coverage by beads, bead type ratio $r=N_\text{M} / N_\text{N}$ with $N_\text{M,N}$ the number of magnetic (M) and non-magnetic (N) beads, and relative susceptibility $\chi_\text{FF}/\chi_\text{M}$, for which an image was recorded, is stated above each column of images. Panel (a) shows an image recorded for large values of $\chi_\text{FF}/\chi_\text{M}$. In this case the interaction between all beads is repulsive and they separate well. The blue beads (magnetic) are slightly closer to each other, as the force between them is smaller compared to the non-magnetic ones. No clustering or percolation is visible. Panel (b) was extracted for equal bead ratio and $\chi_\text{FF}/\chi_\text{M} = 0.5$, resulting in the strongest attractive force between magnetic and non-magnetic beads. Two different beads have a strong tendency to stick together and tile. As the bead ratio is one, they form clusters with four-fold symmetry. Panel (c) shows an image taken at a very unequal bead ratio and $\chi_\text{FF}/\chi_\text{M} = 0.33$. Beads of different kinds attract each other and organize in flower-like structures. However, the excess of one particle species can not be fully compensated, leading to repulsion and many isolated beads. Panel (d) shows the case of unequal bead ratio of 1/4 and $\chi_\text{FF}/\chi_\text{M} = 0.25$. As in panel (c), beads with a large effective moment are coordinated by the beads with smaller moment. However, in this case the excess of one bead species is smaller and in a second step more beads with large dipole moments, attach to the shell of the cluster. As a result larger clusters are formed with local six fold symmetry but still several isolated beads are visible. Panel (e) shows a picture taken with small $\chi_\text{FF}/\chi_\text{M}$ and a bead ratio closer to one. Here it is possible to compensate for the differences in moment and the repulsion between clusters is overcome, forming networks or percolated structures. One might say that one type of beads is the glue keeping the other type, with repulsive interaction, together. The lower panels in Fig.~\ref{fig:imagetable_comp15k} show computer simulations performed for identical parameters as in the upper panels. The simulations look similar to the microscope images, with slightly larger density fluctuations being present in the experimental data. Considering the good visual agreement we have performed more computer simulations and systematically varied all parameters, $\rho$, $r$ and $\chi_\text{FF}/\chi_\text{M}$.

\begin{figure}[t!]
	\centering
		\includegraphics[width=\columnwidth]{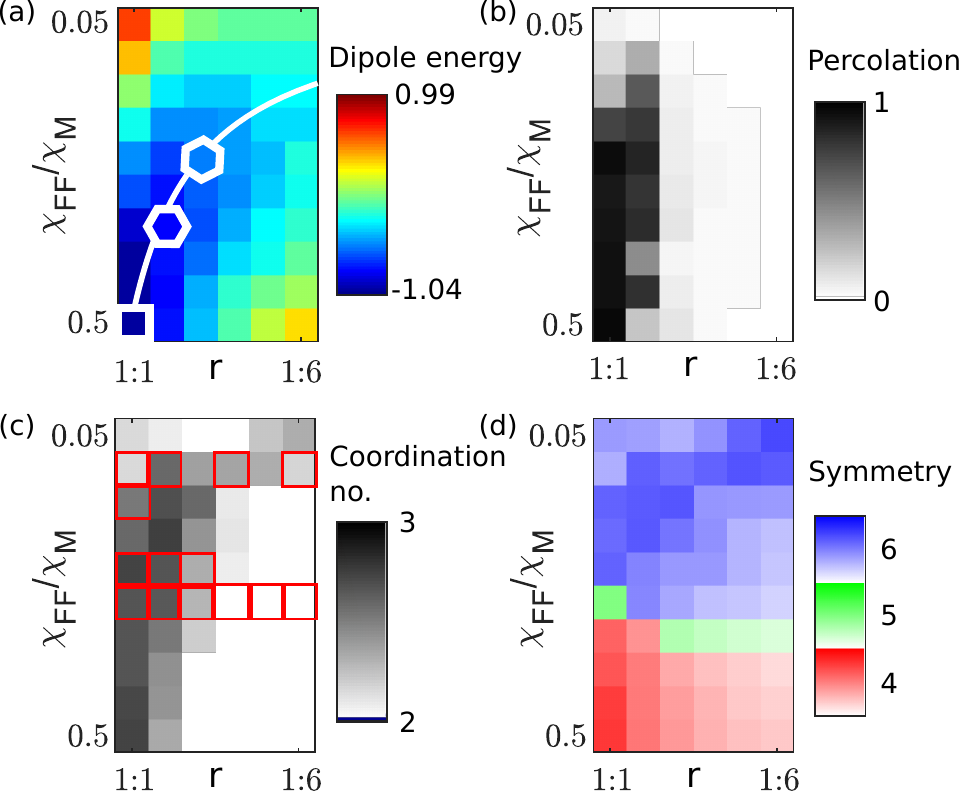}
	\caption{Results from computer simulations performed for a bead concentration of 38.2 \%. Panel (a): Dipole energy per bead plotted versus bead ratio and relative susceptibility. The white line indicates the region of lowest energy and the symbols mark lowest energies for square ($r=1:1$), honeycomb ($r=1:2$) and kagome ($r=1:3$) lattice. Panels (b), (c) and (d) represent the order parameter for percolation, the coordination of individual beads, quantified by the bond order parameter, as well as the most prominent symmetry, respectively. The red squares indicate parameters for which experimental data is available.}
	\label{fig:ratios}
\end{figure}

\section{Discussion}

Fig.~\ref{fig:ratios}(a) depicts the dipole energy per bead plotted as a color map over the bead ratio and relative susceptibility, for a bead density of 38.2\%. This density was chosen as it is close to the critical concentration for percolation as explained later. Negative energies are regions in which the particles have a tendency to form clusters, while in the areas of positive dipole energies, they are isolated. The minimization of dipole energy results from competing attractive and repulsive interactions. Beads of identical magnetic susceptibility experience a repulsive force and the distance between them maximises to minimise the free energy. At the same time, magnetic and non-magnetic beads attract each other, compensating their collective moment and form an in-plane dipole giving rise to an anisotropic short-ranged force field in the plane (decaying faster than $r^{-4}$). The region of lowest energy is indicated by the white line and the white symbols denote the local minimum energy configuration of the beads, which arrange in square, honeycomb and kagome lattices for bead ratios $r$ of 1:1, 1:2 and 1:3, respectively. The coordination number of the beads is shown in Fig.~\ref{fig:ratios}(c). In a chain each bead has two and for hexagonal packing six neighbors. For crystallisation or percolation an average coordination number of at least two is required and open networks may preferentially form for coordination numbers around three. A clear correlation is seen, between larger coordination numbers and low dipole energies per particle, being in line with the formation of large clusters for low dipole energies.

To analyse the coordination of individual beads in more detail,  Fig.~\ref{fig:ratios}(d) shows the symmetry around beads. Blue, green and red colors indicate regions in which predominantly 6-fold, 5-fold and 4-fold symmetry is found, respectively. The intensity of the color represents how often the respective symmetry is found and quantified by the bond order parameter \cite{pham2016crystallization} for a $s$-fold symmetry, defined as $\phi_{j}=\frac{1}{n}\sum \exp(s \cdot i \cdot \theta_k)$, for each bead $j$ having 3 or more neighbors, where $\theta_k$ are the angles of the connecting lines to direct neighbors, with respect to a reference vector. For beads with less than 3 neighbors, the parameter is zero ($\phi_{j}=0$). The average over all beads within one image is $\Phi_s=<|\phi_{j}|>$  and is presented in the colormap. For the apparent moments of the beads being almost equal but of opposite direction four fold symmetries are found. For highly asymmetric moments hexagonal packing is dominant. Interestingly the symmetry is almost independent of the bead ratio, as individual beads, which are left over, remain isolated. This relates well to the fact that large bond order parameters are found in regions of low average dipole moments. Another interesting observation is the five fold symmetry found in the intermediate region. This symmetry hinders the formation of large crystals and should promote the formation of open networks and percolated structures. Fig.~\ref{fig:ratios}(b) depicts the percolation plotted as a map with respect to the bead ratio and relative susceptibility. To extract a percolation order parameter the average cluster size is weighted by the number of beads in each cluster, to become statistically more robust and is normalized to the total number of beads. This quantity can be used as order parameter and varies between zero (isolated beads) and one (all beads connected). Clearly, highly asymmetric bead ratios do not form percolated structures, which are predominantly found for bead ratios of 1:1 and 1:2.

\begin{figure}[t!]
	\centering
		\includegraphics[width=\columnwidth]{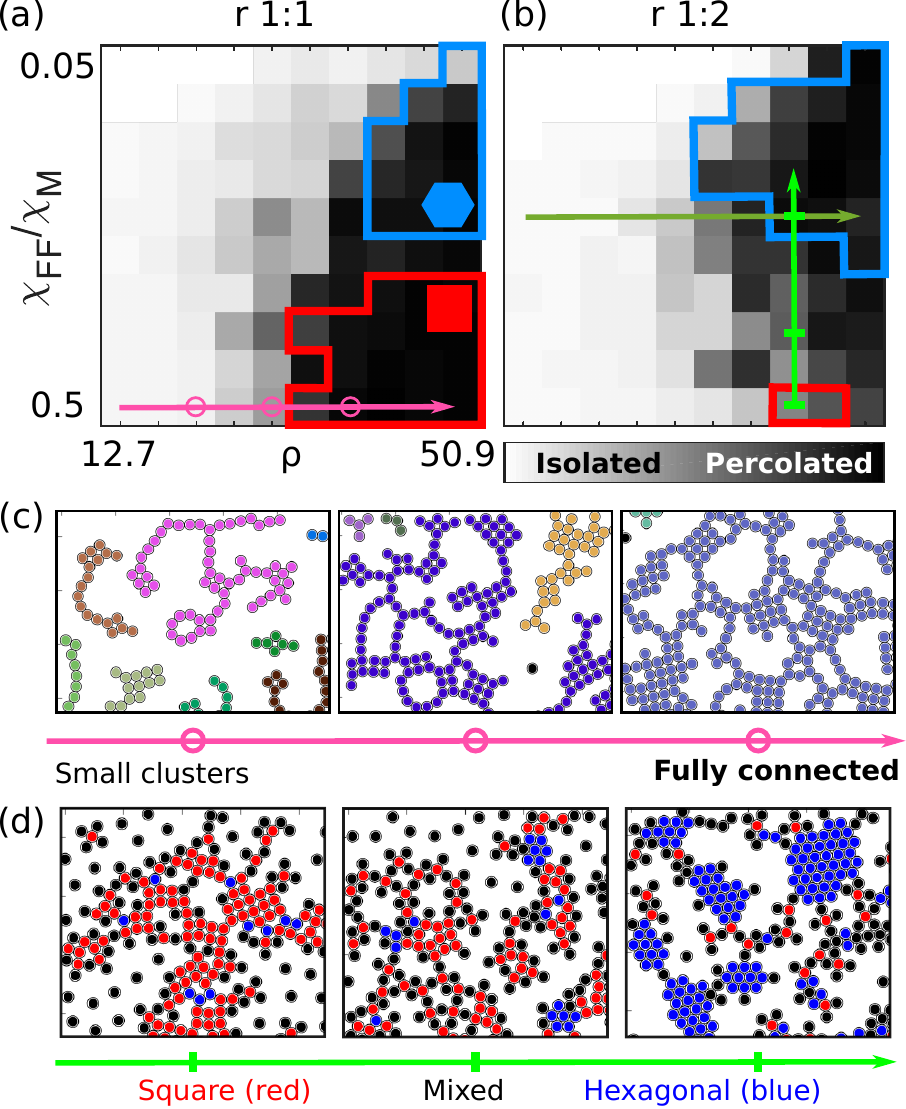}
	\caption{The percolated, hexagonal and square phases plotted over interaction and density for ratios $r=1:1$ (a) and r=1:2 (b). (c)~The transition from isolated beads to percolated with increasing bead density for $r=1:1$. (d)~The transition from square to hexagonal configuration with varying interaction for $r=1:2$.}
	\label{fig:transitions}
\end{figure}

\begin{figure}[t!]
	\centering
		\includegraphics[width=\columnwidth]{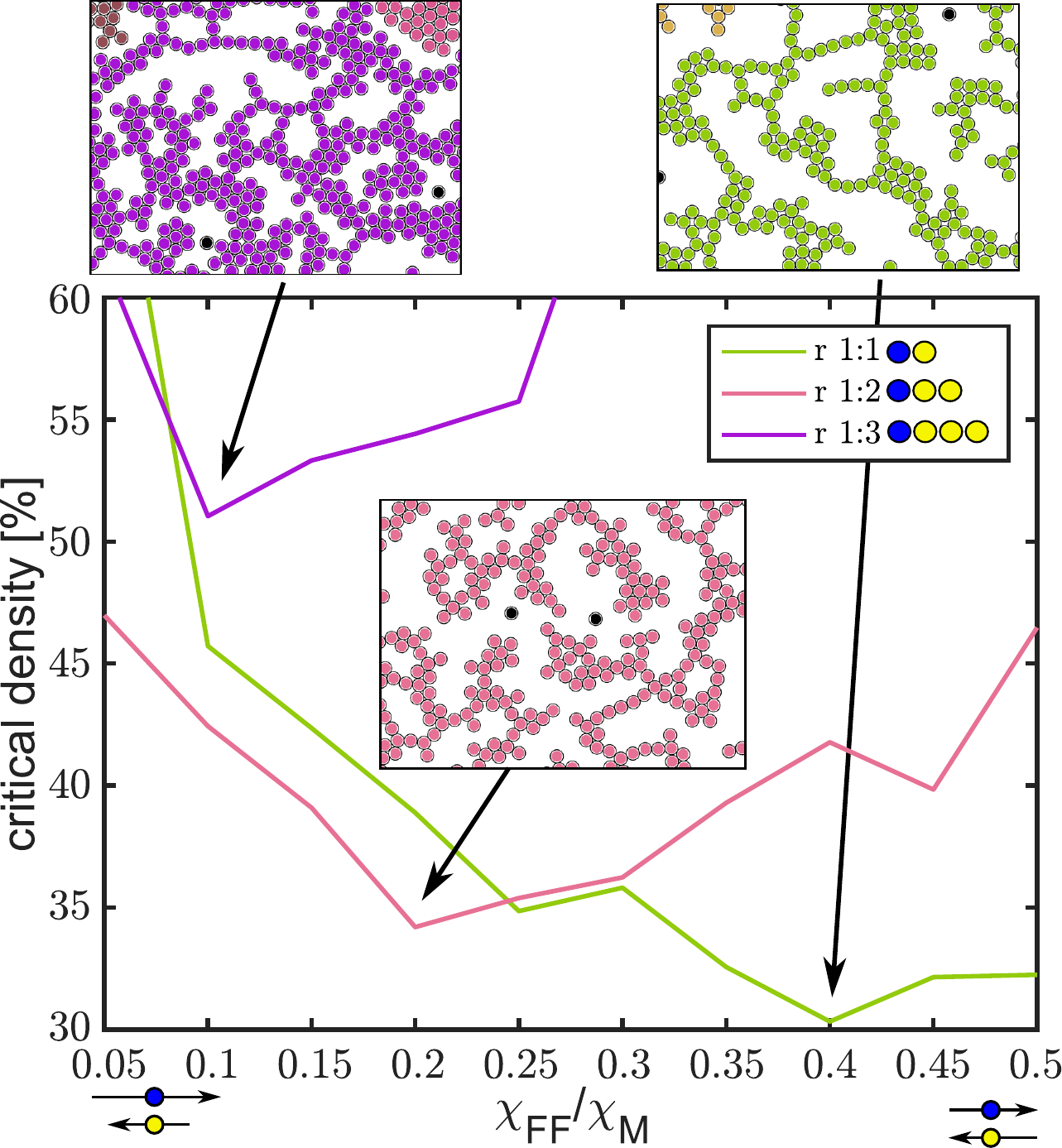}
	\caption{Critical densities for percolation plotted versus interaction of magnetic and non-magnetic beads for bead ratios of 1:1, 1:2 and 1:3.}
		\label{fig5}
\end{figure}

Fig. \ref{fig:ratios} presents results for a bead density of 38.2\% but data has been evaluated for bead densities between 12.7\% and 50.9\%. Fig.~\ref{fig:transitions} depicts the percolation order parameter plotted on a grey-scale versus bead density and interaction, for ratios of 1:1 (panel a) and 1:2 (panel b). The figure was created from simulations. Regions of predominantly hexagonal and cubic symmetry are identified for bond order parameters $\Phi_6$ and $\Phi_4$ larger than 0.4, indicated by the blue and red color respectively.  Fig.~\ref{fig:transitions}(c) depicts simulation images extracted for $\chi_\text{FF}/\chi_\text{M}=0.5$ at three densities, indicated by the purple circles in Fig.~\ref{fig:transitions}(a). For low densities, clusters and/or chains form but do not connect to each other, as their mass is large and their total moment compensated. At higher densities crystallites form and become connected. Fig.~\ref{fig:transitions}(d) depicts simulation images extracted for a density of 38.2 \% and three values of $\chi_\text{FF}/\chi_\text{M}$ indicated by the green bars in Fig.~\ref{fig:transitions}(b). It turns out that in the intermediate region, where neither cubic nor hexagonal symmetries dominate, open percolated structures form, promoted by the large defect density (centre panel, Fig.~\ref{fig:transitions}(d)). The transition to percolated structures with respect to density is relatively sharp for all interaction parameters (Fig.~\ref{fig:transitions}(a) and (b)), allowing the extraction of the critical density for percolation, which is plotted versus $\chi_\text{FF}/\chi_\text{M}$ in Fig.~\ref{fig5}. Bead ratios of 1:1, 1:2 and 1:3 are marked by purple red and green color, respectively.

The critical density for percolation has been calculated for triangles, squares, honeycomb and kagome lattices in two dimensions~\cite{scher1970JCP} and is in all cases close to 45\%. For a bead ratio of 1:1 we find a lower value of the critical density for percolation, of about 30\%, for $\chi_\text{FF}/\chi_\text{M}=0.5$  (top right panel, Fig.~\ref{fig5}). Here the interaction between the two species of beads is attractive and one type acts as glue for the other. The fact that the lowest critical density for percolation is found for values of $\chi_\text{FF}/\chi_\text{M}<0.5$ may be related to crystal tilling effects. At $\chi_\text{FF}/\chi_\text{m}= 0.5$ large cubic ($\rho \approx 79$\%) crystals form leading to larger critical densities for percolation, as it critically depends on the links between crystals. At small values of $\chi_\text{FF}/\chi_\text{M}$, the critical density for percolation becomes large ($>60\%$), as the repulsion between magnetic beads is dominating. A qualitative similar behaviour is found for a bead ratio of 1:2, however, with the lowest value of the critical density for percolation found at lower $\chi_\text{FF}/\chi_\text{M}$. For bead ratios of 1:3 percolation is only found for densities larger than 50\%. The strong repulsion between the magnetic beads requires three non-magnetic beads to compensate, favouring hexagonal dense packed ($\rho>90\%$) arrangements.

\section{Conclusion}

In conclusion, we presented a statistical approach in the analysis of microscopy images and computer simulations, for extracting information on the self-assembly in a two dimensional binary colloidal dispersion with tunable interactions. Magnetic and non-magnetic beads are dissolved in a ferrofluid and investigated in an out of plane magnetic field. Depending on the concentration of magnetic nano-particles in the matrix the interaction between the two types of beads is repulsive or attractive. We find a low critical density for percolation, for attractive forces between the beads and bead ratios approximately equal to one, indicating formation of extended and open networks. Our results are of high relevance for a range of research and technological fields, spanning across physics, chemistry, materials science, biology, epidemics and even large-scale integration of nano-electronic devices \cite{Percolation_Grimmett}. Material systems with controllable interactions, like the one presented here, can serve as an excellent model system as they allow for simple imaging of their microstates, providing statistically relevant information on the self-assembly and formation of percolating structures.

\section*{Author Contributions}
MW, VK and HC together developed the conceptualization and methodology of the project. The investigations and formal analysis were done by HC and AK. HC also developed the software, validated and visualised the results. The project was supervised by MW and VK, who also acquired the funding and administered the project. The original draft was written by HC and then edited by MW and VK and reviewed by all authors.

\section*{Conflicts of interest}
There are no conflicts to declare.

\section*{Acknowledgements}
We thank Niklas Johansson and Anders Olsson, for constructing parts of the experimental setup.
We acknowledge financial support from STINT (contract number: IG2011-2067), Swedish research council (contract number: A0505501), the Carl Tryggers Stiftelse (Contract number: CT 13:513) as well as ABB group.



\balance



\begin{mcitethebibliography}{46}
\providecommand*{\natexlab}[1]{#1}
\providecommand*{\mciteSetBstSublistMode}[1]{}
\providecommand*{\mciteSetBstMaxWidthForm}[2]{}
\providecommand*{\mciteBstWouldAddEndPuncttrue}
  {\def\EndOfBibitem{\unskip.}}
\providecommand*{\mciteBstWouldAddEndPunctfalse}
  {\let\EndOfBibitem\relax}
\providecommand*{\mciteSetBstMidEndSepPunct}[3]{}
\providecommand*{\mciteSetBstSublistLabelBeginEnd}[3]{}
\providecommand*{\EndOfBibitem}{}
\mciteSetBstSublistMode{f}
\mciteSetBstMaxWidthForm{subitem}
{(\emph{\alph{mcitesubitemcount}})}
\mciteSetBstSublistLabelBeginEnd{\mcitemaxwidthsubitemform\space}
{\relax}{\relax}

\bibitem[Whitesides and Grzybowski(2002)]{whitesides2002SCIENCE}
G.~M. Whitesides and B.~Grzybowski, \emph{Science}, 2002, \textbf{295},
  2418\relax
\mciteBstWouldAddEndPuncttrue
\mciteSetBstMidEndSepPunct{\mcitedefaultmidpunct}
{\mcitedefaultendpunct}{\mcitedefaultseppunct}\relax
\EndOfBibitem
\bibitem[Whitesides and Boncheva(2002)]{whitesides2002PNAS}
G.~M. Whitesides and M.~Boncheva, \emph{PNAS}, 2002, \textbf{99}, 4769\relax
\mciteBstWouldAddEndPuncttrue
\mciteSetBstMidEndSepPunct{\mcitedefaultmidpunct}
{\mcitedefaultendpunct}{\mcitedefaultseppunct}\relax
\EndOfBibitem
\bibitem[Klokkenburg \emph{et~al.}(2006)Klokkenburg, Dullens, Kegel, Ern{\'e},
  and Philipse]{klokkenburg2006quantitative}
M.~Klokkenburg, R.~P. Dullens, W.~K. Kegel, B.~H. Ern{\'e} and A.~P. Philipse,
  \emph{Phys. Rev. Lett.}, 2006, \textbf{96}, 037203\relax
\mciteBstWouldAddEndPuncttrue
\mciteSetBstMidEndSepPunct{\mcitedefaultmidpunct}
{\mcitedefaultendpunct}{\mcitedefaultseppunct}\relax
\EndOfBibitem
\bibitem[Qiu \emph{et~al.}(2007)Qiu, Cosgrove, and Howe]{Steric_Langmuir_2007}
D.~Qiu, T.~Cosgrove and A.~M. Howe, \emph{Langmuir}, 2007, \textbf{23},
  475--481\relax
\mciteBstWouldAddEndPuncttrue
\mciteSetBstMidEndSepPunct{\mcitedefaultmidpunct}
{\mcitedefaultendpunct}{\mcitedefaultseppunct}\relax
\EndOfBibitem
\bibitem[Cademartiri \emph{et~al.}(2012)Cademartiri, Stan, Tran, Wu, Friar,
  Vulis, Clark, Tricard, and Whitesides]{C2SM26192H}
R.~Cademartiri, C.~A. Stan, V.~M. Tran, E.~Wu, L.~Friar, D.~Vulis, L.~W. Clark,
  S.~Tricard and G.~M. Whitesides, \emph{Soft Matter}, 2012, \textbf{8},
  9771--9791\relax
\mciteBstWouldAddEndPuncttrue
\mciteSetBstMidEndSepPunct{\mcitedefaultmidpunct}
{\mcitedefaultendpunct}{\mcitedefaultseppunct}\relax
\EndOfBibitem
\bibitem[Demir{\"o}rs \emph{et~al.}(2010)Demir{\"o}rs, Johnson, van Kats, van
  Blaaderen, and Imhof]{demirors2010directed}
A.~F. Demir{\"o}rs, P.~M. Johnson, C.~M. van Kats, A.~van Blaaderen and
  A.~Imhof, \emph{Langmuir}, 2010, \textbf{26}, 14466--14471\relax
\mciteBstWouldAddEndPuncttrue
\mciteSetBstMidEndSepPunct{\mcitedefaultmidpunct}
{\mcitedefaultendpunct}{\mcitedefaultseppunct}\relax
\EndOfBibitem
\bibitem[Kitaev and Ozin(2003)]{kitaev2003self}
V.~Kitaev and G.~A. Ozin, \emph{Adv. Mater.}, 2003, \textbf{15}, 75--78\relax
\mciteBstWouldAddEndPuncttrue
\mciteSetBstMidEndSepPunct{\mcitedefaultmidpunct}
{\mcitedefaultendpunct}{\mcitedefaultseppunct}\relax
\EndOfBibitem
\bibitem[Du \emph{et~al.}(2013)Du, Li, Thakur, and Biswal]{du2013generating}
D.~Du, D.~Li, M.~Thakur and S.~L. Biswal, \emph{Soft Matter}, 2013, \textbf{9},
  6867--6875\relax
\mciteBstWouldAddEndPuncttrue
\mciteSetBstMidEndSepPunct{\mcitedefaultmidpunct}
{\mcitedefaultendpunct}{\mcitedefaultseppunct}\relax
\EndOfBibitem
\bibitem[Kantorovich \emph{et~al.}(2008)Kantorovich, Cerda, and
  Holm]{Kantorovich2008}
S.~Kantorovich, J.~J. Cerda and C.~Holm, \emph{Phys. Chem. Chem. Phys.}, 2008,
  \textbf{10}, 1883\relax
\mciteBstWouldAddEndPuncttrue
\mciteSetBstMidEndSepPunct{\mcitedefaultmidpunct}
{\mcitedefaultendpunct}{\mcitedefaultseppunct}\relax
\EndOfBibitem
\bibitem[Butter \emph{et~al.}(2003)Butter, Bomans, Frederik, Vroege, and
  Philipse]{Butter2003}
K.~Butter, P.~H.~H. Bomans, P.~M. Frederik, G.~J. Vroege and A.~P. Philipse,
  \emph{Nature materials}, 2003, \textbf{2}, 88\relax
\mciteBstWouldAddEndPuncttrue
\mciteSetBstMidEndSepPunct{\mcitedefaultmidpunct}
{\mcitedefaultendpunct}{\mcitedefaultseppunct}\relax
\EndOfBibitem
\bibitem[Weis and Levesque(1993)]{PhysRevLett.71.2729}
J.~J. Weis and D.~Levesque, \emph{Phys. Rev. Lett.}, 1993, \textbf{71},
  2729--2732\relax
\mciteBstWouldAddEndPuncttrue
\mciteSetBstMidEndSepPunct{\mcitedefaultmidpunct}
{\mcitedefaultendpunct}{\mcitedefaultseppunct}\relax
\EndOfBibitem
\bibitem[Kantorovich \emph{et~al.}(2015)Kantorovich, Ivanov, Rovigatti,
  Tavares, and Sciortinoe]{Kantorovich2015}
S.~Kantorovich, A.~O. Ivanov, L.~Rovigatti, J.~M. Tavares and F.~Sciortinoe,
  \emph{Phys. Chem. Chem. Phys.}, 2015, \textbf{17}, 16601\relax
\mciteBstWouldAddEndPuncttrue
\mciteSetBstMidEndSepPunct{\mcitedefaultmidpunct}
{\mcitedefaultendpunct}{\mcitedefaultseppunct}\relax
\EndOfBibitem
\bibitem[Wen \emph{et~al.}(1999)Wen, Kun, P\'al, Zheng, and
  Tu]{PhysRevE.59.R4758}
W.~Wen, F.~Kun, K.~F. P\'al, D.~W. Zheng and K.~N. Tu, \emph{Phys. Rev. E},
  1999, \textbf{59}, R4758--R4761\relax
\mciteBstWouldAddEndPuncttrue
\mciteSetBstMidEndSepPunct{\mcitedefaultmidpunct}
{\mcitedefaultendpunct}{\mcitedefaultseppunct}\relax
\EndOfBibitem
\bibitem[Stambaugh \emph{et~al.}(2003)Stambaugh, Lathrop, Ott, and
  Losert]{PhysRevE.68.026207}
J.~Stambaugh, D.~P. Lathrop, E.~Ott and W.~Losert, \emph{Phys. Rev. E}, 2003,
  \textbf{68}, 026207\relax
\mciteBstWouldAddEndPuncttrue
\mciteSetBstMidEndSepPunct{\mcitedefaultmidpunct}
{\mcitedefaultendpunct}{\mcitedefaultseppunct}\relax
\EndOfBibitem
\bibitem[Stambaugh \emph{et~al.}(2004)Stambaugh, Smith, Ott, and
  Losert]{PhysRevE.70.031304}
J.~Stambaugh, Z.~Smith, E.~Ott and W.~Losert, \emph{Phys. Rev. E}, 2004,
  \textbf{70}, 031304\relax
\mciteBstWouldAddEndPuncttrue
\mciteSetBstMidEndSepPunct{\mcitedefaultmidpunct}
{\mcitedefaultendpunct}{\mcitedefaultseppunct}\relax
\EndOfBibitem
\bibitem[Blair and Kudrolli(2003)]{PhysRevE.67.021302}
D.~L. Blair and A.~Kudrolli, \emph{Phys. Rev. E}, 2003, \textbf{67},
  021302\relax
\mciteBstWouldAddEndPuncttrue
\mciteSetBstMidEndSepPunct{\mcitedefaultmidpunct}
{\mcitedefaultendpunct}{\mcitedefaultseppunct}\relax
\EndOfBibitem
\bibitem[Kr{"o}gel \emph{et~al.}(2018)Kr{"o}gel, Sanchez, Maretzki, Dumont,
  Pyanzina, Kantorovich, and Richter]{Kroger2018}
A.~Kr{"o}gel, P.~A. Sanchez, R.~Maretzki, T.~Dumont, E.~S. Pyanzina, S.~S.
  Kantorovich and R.~Richter, \emph{Soft Matter}, 2018, \textbf{14}, 1001\relax
\mciteBstWouldAddEndPuncttrue
\mciteSetBstMidEndSepPunct{\mcitedefaultmidpunct}
{\mcitedefaultendpunct}{\mcitedefaultseppunct}\relax
\EndOfBibitem
\bibitem[Pham \emph{et~al.}(2017)Pham, Zhuang, Detwiler, Socolar, Charbonneau,
  and Yellen]{Pham2017}
A.~T. Pham, Y.~Zhuang, P.~Detwiler, J.~E.~S. Socolar, P.~Charbonneau and B.~B.
  Yellen, \emph{Phys. Rev. E}, 2017, \textbf{95}, 052607\relax
\mciteBstWouldAddEndPuncttrue
\mciteSetBstMidEndSepPunct{\mcitedefaultmidpunct}
{\mcitedefaultendpunct}{\mcitedefaultseppunct}\relax
\EndOfBibitem
\bibitem[Alert \emph{et~al.}(2014)Alert, Casademunt, and
  Tierno]{alert2014landscape}
R.~Alert, J.~Casademunt and P.~Tierno, \emph{Phys. Rev. Lett.}, 2014,
  \textbf{113}, 198301\relax
\mciteBstWouldAddEndPuncttrue
\mciteSetBstMidEndSepPunct{\mcitedefaultmidpunct}
{\mcitedefaultendpunct}{\mcitedefaultseppunct}\relax
\EndOfBibitem
\bibitem[Straube and Tierno(2014)]{C4SM00132J}
A.~V. Straube and P.~Tierno, \emph{Soft Matter}, 2014, \textbf{10},
  3915--3925\relax
\mciteBstWouldAddEndPuncttrue
\mciteSetBstMidEndSepPunct{\mcitedefaultmidpunct}
{\mcitedefaultendpunct}{\mcitedefaultseppunct}\relax
\EndOfBibitem
\bibitem[Snezhko \emph{et~al.}(2009)Snezhko, Belkin, Aranson, and
  Kwok]{snezhko2009self}
A.~Snezhko, M.~Belkin, I.~Aranson and W.~Kwok, \emph{Phys. Rev. Lett.}, 2009,
  \textbf{102}, 118103\relax
\mciteBstWouldAddEndPuncttrue
\mciteSetBstMidEndSepPunct{\mcitedefaultmidpunct}
{\mcitedefaultendpunct}{\mcitedefaultseppunct}\relax
\EndOfBibitem
\bibitem[Erb \emph{et~al.}(2009)Erb, Son, Samanta, Rotello, and
  Yellen]{erb2009magnetic}
R.~Erb, H.~Son, B.~Samanta, V.~Rotello and B.~Yellen, \emph{Nature}, 2009,
  \textbf{457}, 999--1002\relax
\mciteBstWouldAddEndPuncttrue
\mciteSetBstMidEndSepPunct{\mcitedefaultmidpunct}
{\mcitedefaultendpunct}{\mcitedefaultseppunct}\relax
\EndOfBibitem
\bibitem[Khalil \emph{et~al.}(2012)Khalil, Sagastegui, Li, Tahir, Socolar,
  Wiley, and Yellen]{khalil2012binary}
K.~Khalil, A.~Sagastegui, Y.~Li, M.~Tahir, J.~Socolar, B.~Wiley and B.~Yellen,
  \emph{Nat. Commun.}, 2012, \textbf{3}, 794\relax
\mciteBstWouldAddEndPuncttrue
\mciteSetBstMidEndSepPunct{\mcitedefaultmidpunct}
{\mcitedefaultendpunct}{\mcitedefaultseppunct}\relax
\EndOfBibitem
\bibitem[Carstensen \emph{et~al.}(2015)Carstensen, Kapaklis, and
  Wolff]{carstensen2015phase}
H.~Carstensen, V.~Kapaklis and M.~Wolff, \emph{Phys. Rev. E}, 2015,
  \textbf{92}, 012303\relax
\mciteBstWouldAddEndPuncttrue
\mciteSetBstMidEndSepPunct{\mcitedefaultmidpunct}
{\mcitedefaultendpunct}{\mcitedefaultseppunct}\relax
\EndOfBibitem
\bibitem[Rosensweig \emph{et~al.}(2008)Rosensweig, Hirota, Tsuda, and
  Raj]{rosensweig2008study}
R.~Rosensweig, Y.~Hirota, S.~Tsuda and K.~Raj, \emph{Journal of Physics:
  Condensed Matter}, 2008, \textbf{20}, 204147\relax
\mciteBstWouldAddEndPuncttrue
\mciteSetBstMidEndSepPunct{\mcitedefaultmidpunct}
{\mcitedefaultendpunct}{\mcitedefaultseppunct}\relax
\EndOfBibitem
\bibitem[He \emph{et~al.}(2010)He, Hu, Kim, Ge, Kwon, and Yin]{photonic}
L.~He, Y.~Hu, H.~Kim, J.~Ge, S.~Kwon and Y.~Yin, \emph{Nano Lett.}, 2010,
  \textbf{10}, 4708--4714\relax
\mciteBstWouldAddEndPuncttrue
\mciteSetBstMidEndSepPunct{\mcitedefaultmidpunct}
{\mcitedefaultendpunct}{\mcitedefaultseppunct}\relax
\EndOfBibitem
\bibitem[He \emph{et~al.}(2012)He, Wang, Ge, and Yin]{he2012magnetic}
L.~He, M.~Wang, J.~Ge and Y.~Yin, \emph{Acc. Chem. Res.}, 2012, \textbf{45},
  1431--1440\relax
\mciteBstWouldAddEndPuncttrue
\mciteSetBstMidEndSepPunct{\mcitedefaultmidpunct}
{\mcitedefaultendpunct}{\mcitedefaultseppunct}\relax
\EndOfBibitem
\bibitem[Saado \emph{et~al.}(2002)Saado, Golosovsky, Davidov, and
  Frenkel]{saado2002tunable}
Y.~Saado, M.~Golosovsky, D.~Davidov and A.~Frenkel, \emph{Phys. Rev. B}, 2002,
  \textbf{66}, 195108\relax
\mciteBstWouldAddEndPuncttrue
\mciteSetBstMidEndSepPunct{\mcitedefaultmidpunct}
{\mcitedefaultendpunct}{\mcitedefaultseppunct}\relax
\EndOfBibitem
\bibitem[Fan \emph{et~al.}(2013)Fan, Chen, Lin, Miao, Chang, Liu, Wang, and
  Lin]{fan2013magnetically}
F.~Fan, S.~Chen, W.~Lin, Y.-P. Miao, S.-J. Chang, B.~Liu, X.-H. Wang and
  L.~Lin, \emph{Appl. Phys. Lett.}, 2013, \textbf{103}, 161115\relax
\mciteBstWouldAddEndPuncttrue
\mciteSetBstMidEndSepPunct{\mcitedefaultmidpunct}
{\mcitedefaultendpunct}{\mcitedefaultseppunct}\relax
\EndOfBibitem
\bibitem[Nguyen and Choi(2009)]{nguyen2009optimal}
Q.-H. Nguyen and S.-B. Choi, \emph{Smart Mater. Struct.}, 2009, \textbf{18},
  035012\relax
\mciteBstWouldAddEndPuncttrue
\mciteSetBstMidEndSepPunct{\mcitedefaultmidpunct}
{\mcitedefaultendpunct}{\mcitedefaultseppunct}\relax
\EndOfBibitem
\bibitem[Park \emph{et~al.}(2010)Park, Fang, and Choi]{C0SM00014K}
B.~J. Park, F.~F. Fang and H.~J. Choi, \emph{Soft Matter}, 2010, \textbf{6},
  5246--5253\relax
\mciteBstWouldAddEndPuncttrue
\mciteSetBstMidEndSepPunct{\mcitedefaultmidpunct}
{\mcitedefaultendpunct}{\mcitedefaultseppunct}\relax
\EndOfBibitem
\bibitem[Wang and Liao(2011)]{wang2011magnetorheological}
D.~Wang and W.~Liao, \emph{Smart Mater. Struct.}, 2011, \textbf{20},
  023001\relax
\mciteBstWouldAddEndPuncttrue
\mciteSetBstMidEndSepPunct{\mcitedefaultmidpunct}
{\mcitedefaultendpunct}{\mcitedefaultseppunct}\relax
\EndOfBibitem
\bibitem[O'Hern \emph{et~al.}(2003)O'Hern, Silbert, Liu, and
  Nagel]{ohern2003PRE}
C.~S. O'Hern, L.~E. Silbert, A.~J. Liu and S.~R. Nagel, \emph{Phys. Rev. E},
  2003, \textbf{68}, 011306\relax
\mciteBstWouldAddEndPuncttrue
\mciteSetBstMidEndSepPunct{\mcitedefaultmidpunct}
{\mcitedefaultendpunct}{\mcitedefaultseppunct}\relax
\EndOfBibitem
\bibitem[Trappe \emph{et~al.}(2001)Trappe, Prasad, Cipelletti, Segre, and
  Weitz]{trappe2001NATURE}
V.~Trappe, V.~Prasad, L.~Cipelletti, P.~N. Segre and D.~A. Weitz,
  \emph{Nature}, 2001, \textbf{411}, 772\relax
\mciteBstWouldAddEndPuncttrue
\mciteSetBstMidEndSepPunct{\mcitedefaultmidpunct}
{\mcitedefaultendpunct}{\mcitedefaultseppunct}\relax
\EndOfBibitem
\bibitem[Drocco \emph{et~al.}(2005)Drocco, Hastings, Reichhardt, and
  Reichhardt]{Charles_2005_PRL}
J.~A. Drocco, M.~B. Hastings, C.~J.~O. Reichhardt and C.~Reichhardt,
  \emph{Phys. Rev. Lett.}, 2005, \textbf{95}, 088001\relax
\mciteBstWouldAddEndPuncttrue
\mciteSetBstMidEndSepPunct{\mcitedefaultmidpunct}
{\mcitedefaultendpunct}{\mcitedefaultseppunct}\relax
\EndOfBibitem
\bibitem[Koeze and Tighe(2018)]{koeze2018PRL}
D.~J. Koeze and B.~P. Tighe, \emph{Phys. Rev. Lett.}, 2018, \textbf{121},
  188002\relax
\mciteBstWouldAddEndPuncttrue
\mciteSetBstMidEndSepPunct{\mcitedefaultmidpunct}
{\mcitedefaultendpunct}{\mcitedefaultseppunct}\relax
\EndOfBibitem
\bibitem[Achlioptas \emph{et~al.}(2009)Achlioptas, D'Souza, and
  Spencer]{achlioptas2009SCIENCE}
D.~Achlioptas, R.~M. D'Souza and J.~Spencer, \emph{Science}, 2009,
  \textbf{323}, 1453\relax
\mciteBstWouldAddEndPuncttrue
\mciteSetBstMidEndSepPunct{\mcitedefaultmidpunct}
{\mcitedefaultendpunct}{\mcitedefaultseppunct}\relax
\EndOfBibitem
\bibitem[Scher and Zallen(1970)]{scher1970JCP}
H.~Scher and R.~Zallen, \emph{J. Chem. Phys.}, 1970, \textbf{53}, 3759\relax
\mciteBstWouldAddEndPuncttrue
\mciteSetBstMidEndSepPunct{\mcitedefaultmidpunct}
{\mcitedefaultendpunct}{\mcitedefaultseppunct}\relax
\EndOfBibitem
\bibitem[Valadez-P{\'e}rez \emph{et~al.}(2013)Valadez-P{\'e}rez,
  Casta{\~n}eda-Priego, and Liu]{valadez2013RSC}
N.~Valadez-P{\'e}rez, R.~Casta{\~n}eda-Priego and Y.~Liu, \emph{RSC Advances},
  2013, \textbf{3}, 25110\relax
\mciteBstWouldAddEndPuncttrue
\mciteSetBstMidEndSepPunct{\mcitedefaultmidpunct}
{\mcitedefaultendpunct}{\mcitedefaultseppunct}\relax
\EndOfBibitem
\bibitem[Skjeltorp(1984)]{skjeltorp1984crystallization}
A.~Skjeltorp, \emph{Physica B+C}, 1984, \textbf{127}, 411--416\relax
\mciteBstWouldAddEndPuncttrue
\mciteSetBstMidEndSepPunct{\mcitedefaultmidpunct}
{\mcitedefaultendpunct}{\mcitedefaultseppunct}\relax
\EndOfBibitem
\bibitem[Carstensen \emph{et~al.}(2018)Carstensen, Kapaklis, and
  Wolff]{carstensen2018statistical}
H.~Carstensen, V.~Kapaklis and M.~Wolff, \emph{The European Physical Journal
  E}, 2018, \textbf{41}, 9\relax
\mciteBstWouldAddEndPuncttrue
\mciteSetBstMidEndSepPunct{\mcitedefaultmidpunct}
{\mcitedefaultendpunct}{\mcitedefaultseppunct}\relax
\EndOfBibitem
\bibitem[Carstensen(2018)]{phdthesisHauke}
H.~Carstensen, \emph{PhD thesis}, Uppsala University, 2018\relax
\mciteBstWouldAddEndPuncttrue
\mciteSetBstMidEndSepPunct{\mcitedefaultmidpunct}
{\mcitedefaultendpunct}{\mcitedefaultseppunct}\relax
\EndOfBibitem
\bibitem[Yang \emph{et~al.}(2013)Yang, Gao, Lopez, and Yellen]{yang2013tunable}
Y.~Yang, L.~Gao, G.~P. Lopez and B.~B. Yellen, \emph{ACS Nano}, 2013,
  \textbf{7}, 2705--2716\relax
\mciteBstWouldAddEndPuncttrue
\mciteSetBstMidEndSepPunct{\mcitedefaultmidpunct}
{\mcitedefaultendpunct}{\mcitedefaultseppunct}\relax
\EndOfBibitem
\bibitem[Meeker(2021)]{FEMM}
D.~Meeker, \emph{Finite Element Method Magnetics ({FEMM})},
  \url{http://www.femm.info/wiki/HomePage}, 2021\relax
\mciteBstWouldAddEndPuncttrue
\mciteSetBstMidEndSepPunct{\mcitedefaultmidpunct}
{\mcitedefaultendpunct}{\mcitedefaultseppunct}\relax
\EndOfBibitem
\bibitem[Pham \emph{et~al.}(2016)Pham, Seto, Sch{\"o}nke, Joh, Chilkoti, Fried,
  and Yellen]{pham2016crystallization}
A.~T. Pham, R.~Seto, J.~Sch{\"o}nke, D.~Y. Joh, A.~Chilkoti, E.~Fried and B.~B.
  Yellen, \emph{Soft Matter}, 2016, \textbf{12}, 7735--7746\relax
\mciteBstWouldAddEndPuncttrue
\mciteSetBstMidEndSepPunct{\mcitedefaultmidpunct}
{\mcitedefaultendpunct}{\mcitedefaultseppunct}\relax
\EndOfBibitem
\bibitem[Grimmett(1989)]{Percolation_Grimmett}
G.~Grimmett, \emph{Percolation}, Springer-Verlag New York, 1989\relax
\mciteBstWouldAddEndPuncttrue
\mciteSetBstMidEndSepPunct{\mcitedefaultmidpunct}
{\mcitedefaultendpunct}{\mcitedefaultseppunct}\relax
\EndOfBibitem
\end{mcitethebibliography}

\providecommand*{\mcitethebibliography}{\thebibliography}
\csname @ifundefined\endcsname{endmcitethebibliography}
{\let\endmcitethebibliography\endthebibliography}{}

\end{document}